\documentclass[showpacs,twocolumn,prl,aps]{revtex4}
\usepackage{graphicx}

\begin{document}

\title[Short title for running header]{The spin - $\frac{1}{2}$ Heisenberg model on Kagome lattice as a quantum critical system}
\author{Tao Li}
\affiliation{ Department of Physics, Renmin University of China,
Beijing 100872, P.R.China}
\date{\today}

\begin{abstract}
Through exact diagonalization study of the spin - $\frac{1}{2}$
Heisenberg model on Kagome lattice with ring-exchange coupling
$J_{r}$, we find the pure Heisenberg model with $J_{r}=0$ stands as
a quantum critical point, as evidenced by avoided level crossing and
divergence of the second derivative of the ground state energy with
respect to $J_{r}$ at $J_{r}=0$. The pure Heisenberg model should
thus be gapless in the thermodynamic limit, contrary to common
beliefs. At the same time, the ring exchange coupling is found to
drive the system into a state with more short ranged spin
correlation and with a local spin correlation pattern equivalent to
that of the antiferromagnetic Heisenberg model on triangular
lattice(the peak of spin structure factor moves to the momentum
$\vec{q}=(\frac{4\pi}{3},0)$). The resemblance of state with the
Marston-Zeng spin Pereils solid state(in terms of dimer-dimer
correlation) is also much enhanced by the ring exchange coupling,
although it is unclear if such correlation would solidify into
static order in the thermodynamic limit.
\end{abstract}
\maketitle

The spin liquid state represents a novel state of matter beyond the
Landau-Ginzburg description and supports new kinds of order and
excitation. It is generally believed that the geometrically
frustrated quantum antiferromagnetic systems are ideal places to
find spin liquid ground state\cite{review}. The antiferromagnetic
Heisenberg model on Kagome lattice, which is a typical frustrated
system, is a especially promising target\cite{elser1}. The geometric
frustration of the Kagome system is so strong that the classical
spin defined on such a lattice posses extensive ground state
entropy. Understanding how quantum fluctuation would lift such huge
degeneracy stands as a big challenge to both condensed matter theory
and our imagination.

Much efforts, both in theory and experiment\cite{real}, has been
devoted to the understanding of its exotic ground state and
excitation properties. Earlier numerical studies on finite size
systems have accumulated strong evidence for a spin disordered
ground state with no symmetry breaking and a very short ranged spin
correlation\cite{elser2,chalker,singh,leung,young,lhuillier,lhuillier3,series,jiang,vidal,white,lauchli2}.
What makes the Kagome system even more extraordinary is the fact the
spin gap is extremely
small\cite{lhuillier,lhuillier3,jiang,white,lauchli2}, although the
spin correlation length is estimated to be only of the order of one
lattice constant. At the same time, the singlet excitation channel
is found to be abundant of low energy excitations below the spin
gap\cite{lhuillier,lhuillier3,waldtmann,lhuillier2,impurity,specific,lauchli1},
whose origin is still
unclear\cite{elser1,marston,elser3,mila,impurity,auerbach,poilblanc}.
All these characteristics are not what one expect for a typical spin
liquid state with short range spin correlation and have puzzled
people in the field for more than two decades.

It is now generally believed that a small but finite spin gap exists
in the thermodynamic
limit\cite{lhuillier,lhuillier2,jiang,white,lauchli2}. The same is
also believed to be true for the excitation gap in the singlet
channel\cite{white,lauchli2}. The ground state of the Heisenberg
model on Kagome lattice is thus believed to be still consistent with
our understanding for a nominal incompressible and featureless
quantum liquid, although the excitation gap is very small for some
unknown reasons. However, since all these results are all derived
from studies on finite size system and the low energy spectral
weight of the system is very susceptible to local
perturbations\cite{impurity} as a result of the remarkable softness
of the spin dynamics of the system\cite{elser2}, it is very
susceptible that the observed excitation gap may simply be the
artifact of the finite size effect or boundary effect. Slave Boson
mean field analysis has predicted a critical state  with Dirac type
spinon dispersion as the ground state\cite{hastings}, which is shown
later by Variational Monte Carlo calculation to be a good
variational state in terms of energy\cite{ran}. However, many
earlier numerical findings, especially the large amount of singlet
excitation spectral weight below the spin gap(on finite size
system), can not find a natural understanding in such a description.

Here we are especially interested in the low energy excitation in
the singlet channel. Taking it literally, the aggregation of the
large amount of spectral weight at such a low energy can be
interpreted as a result of quasi-degeneracy of the ground state,
which also implies the sensitivity of the ground state on small
perturbations. A close analog of this situation can be found in a
fractionally filled lowest Landau level, in which case the Coulomb
repulsion will reorganize the system into an incompressible quantum
liquid- the fractional quantum Hall state. It is very interesting to
see if similar thing can happen in the Kagome system if we turn on
some relevant perturbation.

Along this line of thinking, we suggest to study the spin -
$\frac{1}{2}$ Heisenberg model on Kagome lattice with ring exchange
coupling. The model is given as follows
\begin{equation}
H=J\sum_{<i,j>}\vec{\mathrm{S}}_{i}\cdot\vec{\mathrm{S}}_{j}+J_{r}\sum_{hexagons
}(P+P^{-1}),
\end{equation}
in which $J_{r}$ denotes the ring exchange coupling constant on each
hexagonal plaquette of the Kagome lattice(shown in Fig.1), $P$
denotes the operator for cyclic permutation of the six electron on a
hexagonal ring. It should be noted that the ring exchange coupling
on hexagonal plaquette represents the lowest order correction to the
pure Heisenberg model in the large $U$ expansion of the Hubbard
model. For realistic systems, $J_{r}$ should have a positive sign.
It is the purpose of this paper to determine if such a correction
would constitute a relevant perturbation on the physics of the pure
Heisenberg system. Especially, we want to see if the ring exchange
coupling would paly a similar role as the Coulomb repulsion in
fractionally filled lowest Landau level system and drive the system
into an incompressible quantum liquid that match our general
expectation for a spin liquid with short range spin correlation.

The effect of the ring exchange coupling has been studied
extensively on frustrated quantum magnet. Earlier numerical study
has found on triangular lattice that the ring exchange coupling may
drive the system into a spin liquid state from the three sublattice
magnetic ordered state\cite{triangular}. However, the nature of the
obtained spin liquid state is still under debate. Some author
proposed a gapless spin liquid state with spinon Fermi surface based
on mean filed analysis\cite{fermi} but others believe that some kind
of spin gap exist\cite{dwave}.

To settle down the issue of the effect of the ring exchange coupling
on the Kagome Heisenberg model, we have carried out exact
diagonalization study on a Kagome cluster with 36 sites(shown in
Fig.1). Such a finite cluster has the important property that it
respects all the symmetry of the original Hamiltonian in the
thermodynamic limit. Such symmetries are believed to be crucial to
the understanding of delicate physics of the Kagome Heisenberg
system. For example, numerical study shows that the low energy
singlet dynamics of Kagome Heisenberg system, which stands as its
characteristic exoticness, can be easily quenched by tiny amount of
vacancy in the lattice\cite{impurity}. The structure of the 36-site
cluster is shown in Figure 1. The same cluster is first used in a
exact diagonalization study of the pure Heisenberg model by Leung
and Elser in 1993\cite{leung}. Here we have adopted the same
numbering of sites and bonds as their choice for comparison of
results.

\begin{figure}[h!]
\includegraphics[width=8cm,angle=0]{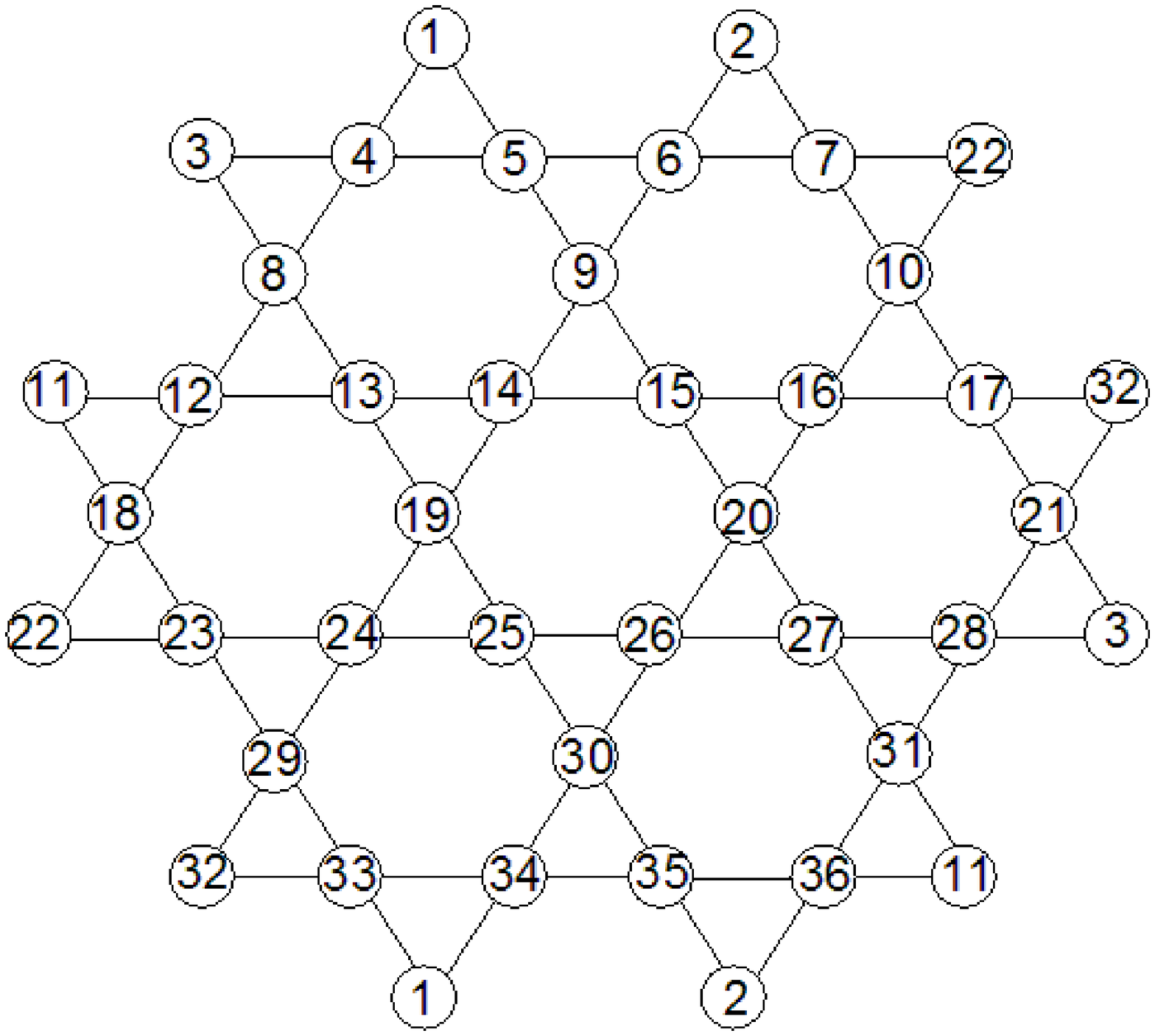}
\includegraphics[width=6cm,angle=0]{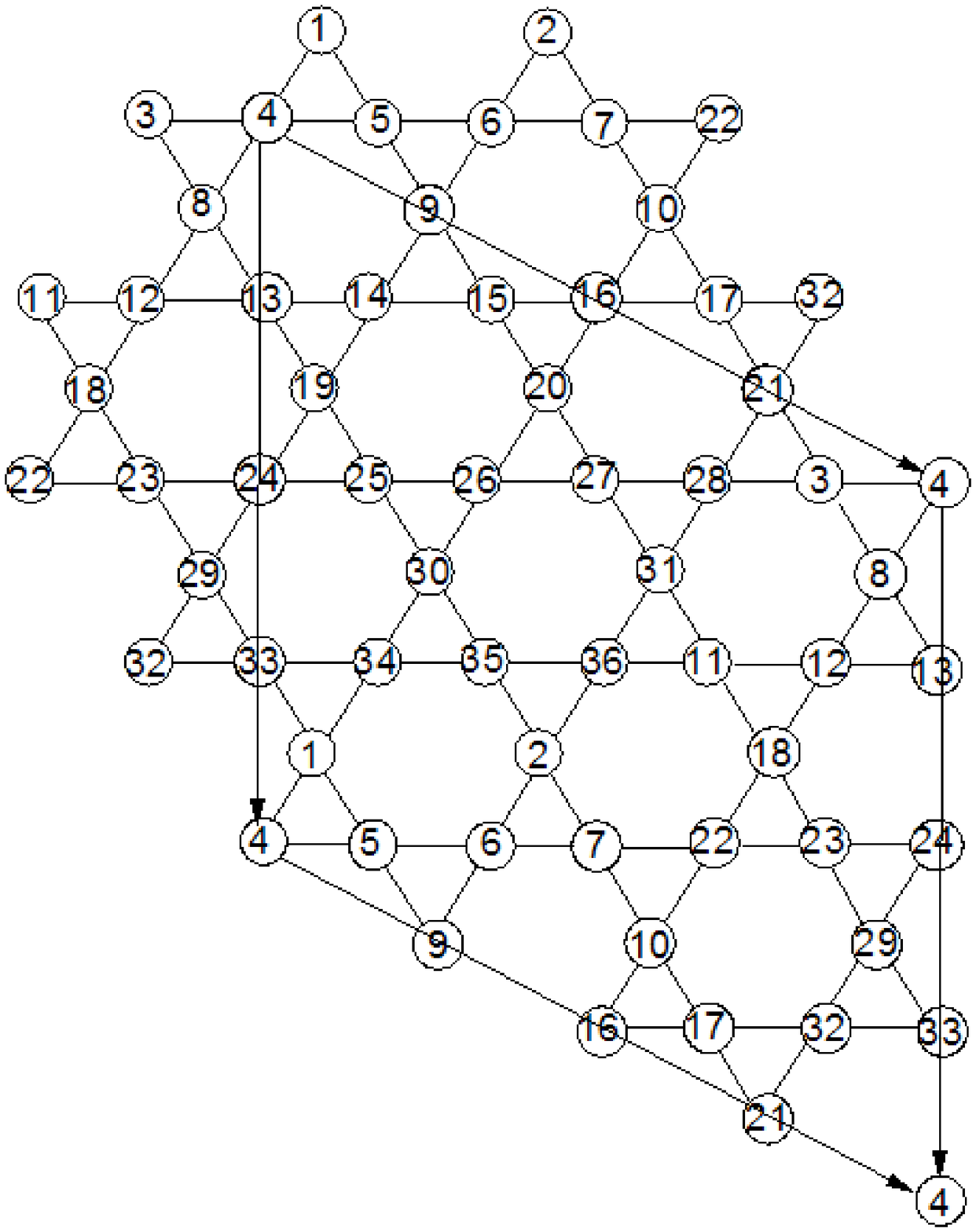}
\caption{The 36-site Kagome cluster on which our exact
diagonalization study is performed. Our numbering of the lattice
sites is the same as that used in Ref\cite{leung}. In the upper
panel, the cluster is presented in such a way that the rotational
symmetry of the cluster is explicit, while in the lower panel, it is
presented in such a fashion that the boundary condition on the
cluster is explicit.} \label{fig1}
\end{figure}

In our study, we have concentrated on the fully symmetric subspace
in which the ground state resides. This subspace, which belongs to
the identity representation of the symmetry group, has 31527894
basis vectors. We have used both the Lanczos and the Arnoldi
algorithm\cite{parpack} to calculate the ground state properties and
the lowest excitation in the fully symmetric subspace. Lanczos
calculation of the lowest eigenvalue in general irreducible
representations at high symmetry momentum is carried out also to
make sure that ground state indeed resides in the fully symmetric
subspace. To uncover the nature of the ground state found, we have
calculated the spin-spin correlation and dimer-dimer correlation
function on the cluster.

Although the physical ring exchange coupling should be positive, we
will treat $J_{r}$ as a free controlling parameter that can take
both positive and negative values. In Figure 2, we plot the ground
state energy and the lowest excitation energy in the fully symmetric
subspace as functions of $J_{r}$. The most striking thing in this
figure is the fact that an avoided level crossing occurs exactly at
the pure Heisenberg limit, namely at $J_{r}=0$.

\begin{figure}[h!]
\includegraphics[width=8cm,angle=0]{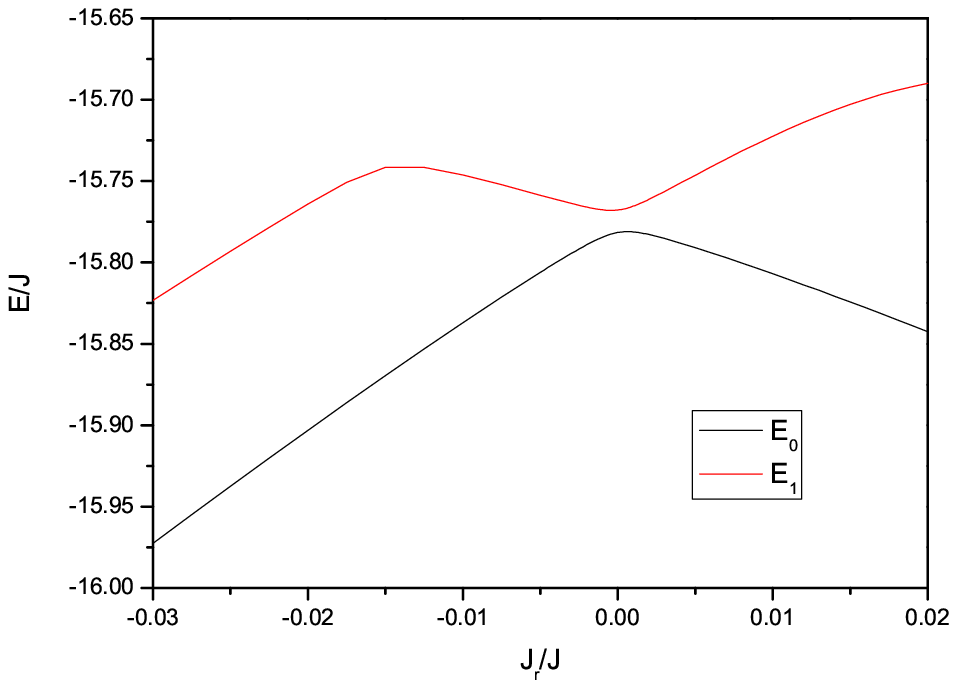}
\includegraphics[width=8cm,angle=0]{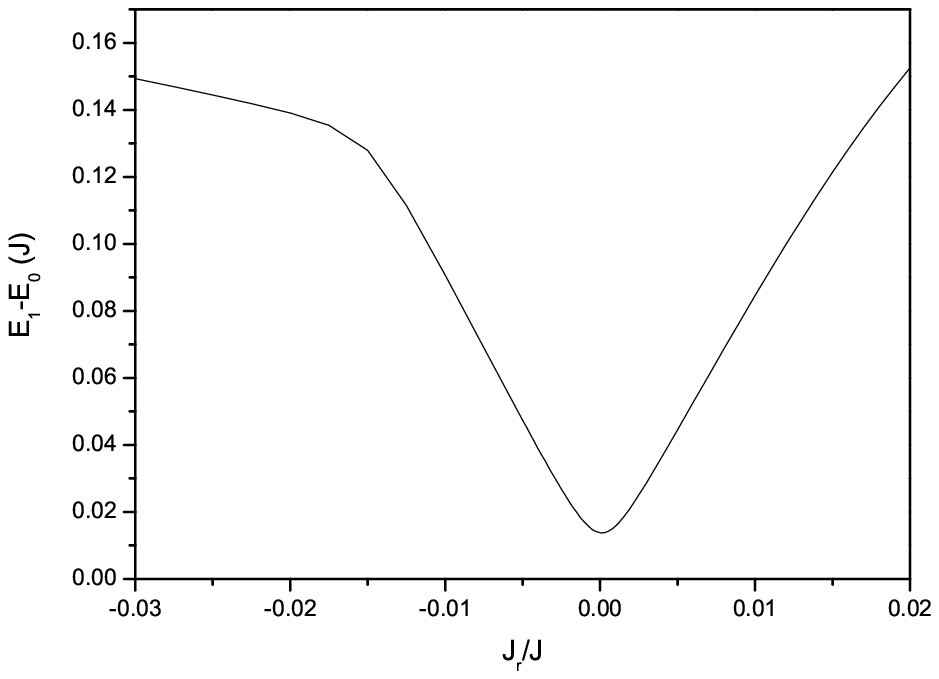}
\caption{Upper panel: The lowest two eigenvalues in the fully
symmetric subspace as functions of the ring exchange coupling.
Avoided level crossing is clearly seen around the pure Heisenberg
limit at $J_{r}=0$. Lower panel: the difference in the lowest two
eigenvalues as a function of
 $J_{r}$.} \label{fig2}
\end{figure}

An avoided level crossing in the spectrum of a finite system is a
signature of the quantum phase transition\cite{sachdev}. As the size
of the lattice grows, the crossing will become progressively sharper
and eventually lead to a non-analyticity in ground state energy in
the infinite lattice limit. Thanks to the short ranged nature of the
spin correlation in the ground state of the pure Heisenberg model,
the $36$-site cluster system already exhibits quite sharp level
crossing. To make it clearer, we plot in Figure 2 also the
difference of the crossing energies as a function of $J_{r}$.

Now we focus on the ground state properties. As the transition is
driven by the ring exchange coupling, we plot the first and the
second derivative of the ground state energy as a function of
$J_{r}$ in Fig.4. According to the Hellmann-Feynman theorem, the
first derivative is just the expectation value of the ring exchange
coupling in the ground state. It is expected that such coupling
should engage in a dramatic change acrossing the transition point at
$J_{r}=0$, as is clear in the figure. The second derivative shown in
Fig. 4(b) exhibits the typical divergence during a quantum phase
transition, although the calculation is done on a still relatively
small lattice of 36 sites. The pure Heisenberg model on Kagome
lattice is thus a quantum critical point. As a result, the gap in
the spin channel and singlet channel should all vanish in the
thermodynamic limit. This is contrary to the common beliefs on these
issues.

\begin{figure}[h!]
\includegraphics[width=7cm,angle=0]{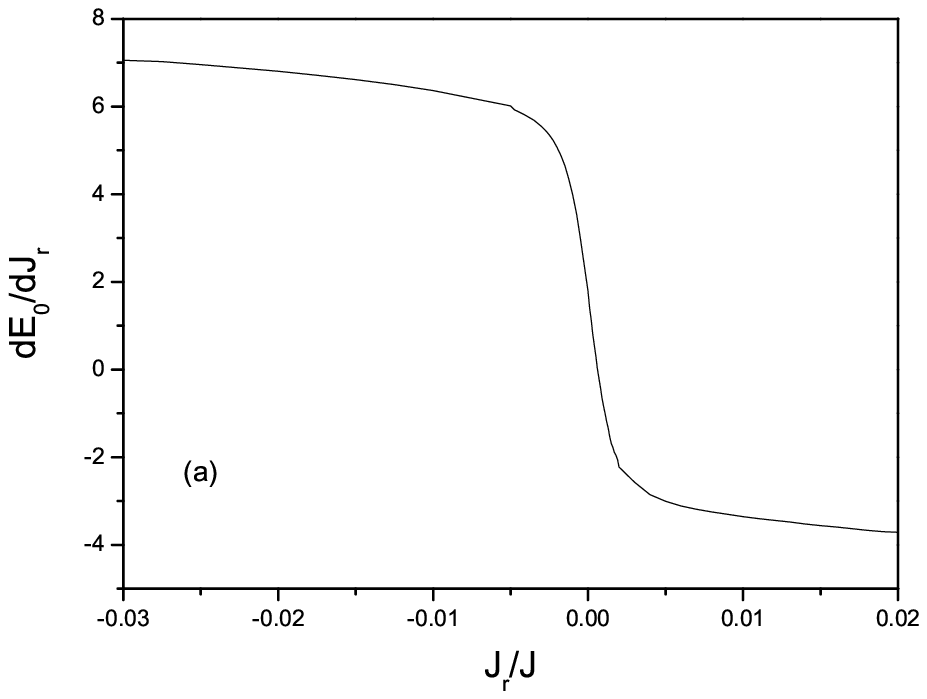}
\includegraphics[width=7cm,angle=0]{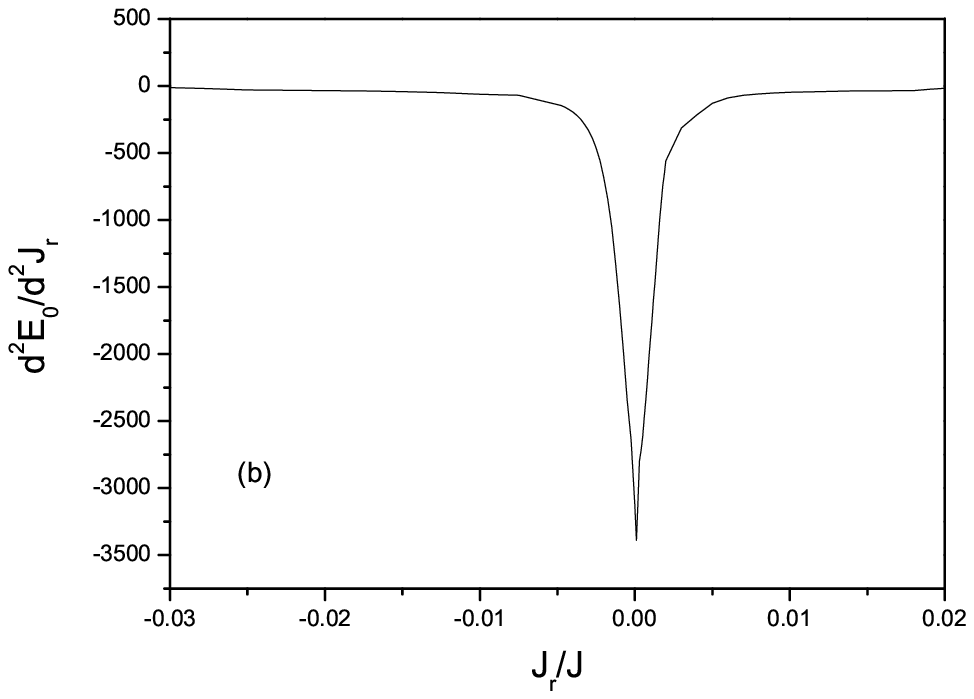}
\caption{The first (a) and the second (b) derivative of the ground
state energy as a function of $J_{r}$.} \label{fig3}
\end{figure}

After establishing that the pure Heisenberg model lies at a quantum
critical point, we now move to the question what state the ring
exchange coupling will drive the system into. For this purpose, we
have examined the spin-spin correlation and dimer-dimer correlation
function. The comparison of the spin correlation function between
the pure Heisenberg mode and the model with $J_{r}/J=0.15$ for all
the 10 inequivalent distances on the 36-site cluster is tabulated in
Table I. The numbering of sites are the same as that used in
Ref\cite{leung} and we also use site $26$ as the reference site. The
main difference between the pure Heisenberg model and the
$J_{r}=0.15J$ system can be summarized as follows. The spin
correlation between spins on the same hexagonal ring(site $14$ and
site $15$) is enhanced by the ring exchange coupling, while those
outside the same hexagonal ring is in general reduced. This results
in a more short-ranged spin correlation function. This is can be
seen more clearly in Fig.5, in which we plot the absolute value of
the spin correlation between the reference site $26$ and all other
sites.

\begin{figure}[h!]
\includegraphics[width=7cm,angle=0]{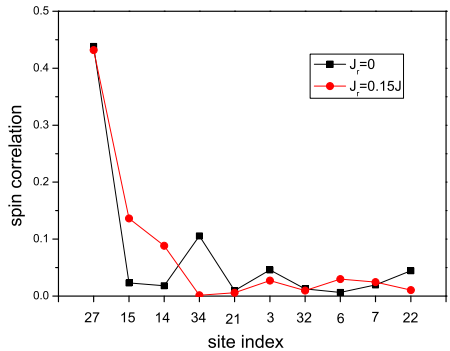}
\caption{The absolute value of the spin correlation function between
the reference site $26$ and all the 10 symmetry inequivalent sites
on the 36-site cluster. The ordering of the sites in the figure is
determined by their distance with site $26$. When such distance is
degenerate(for example, site $14$ and $34$), the site on the same
hexagonal ring as site $26$ appear earlier in the ordering. }
\label{fig4}
\end{figure}

To see if there is any qualitative difference in the spin
correlation between the pure Heisenberg model and the model with
non-zero ring exchange coupling, we have calculated the spin
structure factor at both $J_{r}=0$ and $J_{r}=0.15J$. The spin
structure factor is defined as
\begin{equation}
S(q)=\sum_{i,j}e^{iq\cdot (R_{i}-R_{j})}\langle S_{i}\cdot
S_{j}\rangle.
\end{equation}
On the finite cluster, care should be paid on the choice of the wave
vector $\vec{q}$. In the lower panel of Fig.1, we have plotted the
translational unit of the 36-site cluster, on which periodic
boundary condition is imposed. Under such boundary condition, the
allowed momentum can be generally written as
\begin{equation}
\vec{q}=q_{x}\vec{b}_{1}+q_{y}\vec{b}_{2},
\end{equation}
in which $\vec{b}_{1,2}=2\pi(1,\pm\frac{1}{\sqrt{3}})$ denotes the
two primitive reciprocal vectors of the triangular lattice from
which the Kagome lattice is derived by removing one fourth of
lattice sites. $q_{x}=\frac{m}{12},q_{y}=\frac{n}{12}$ with
$m,n=0,\cdots, 12$. Although there is clearly redundancy in the
momentum mesh Eq.(3) for the description of spatial variation on the
36-site cluster, it presents a natural way to understand the
evolution of the spin correlation pattern on the Kagome lattice, as
will be clear below.

The spin structure factor for the pure Heisenberg model and the
model with $J_{r}=0.15J$ are shown in Fig.6. For the pure Heisenberg
model, the spin structure factor peaks at three independent momentum
$(q_{x},q_{y})=(0,1/2),(1/2,0),(1/2,1/2)$. This structure is caused
by the antiferromagnetic correlation between nearest neighboring
spins in the three directions. For the $J_{r}=0.15J$ case, the peak
of the spin structure factor moves to $(q_{x},q_{y})=(1/3,1/3)$, or
to the momentum $\vec{q}=(\frac{4\pi}{3},0)$. This momentum is
characteristic of the coplanar local spin correlation on triangular
lattice with 120 degree angle between neighboring spins on each
triangular plaquette(it is interesting to note that the 36-site
cluster studied in this paper can host such a momentum). At the same
time, the peak in the spin structure factor is found to be more
rounded than the pure Heisenberg model, indicating that the spin
correlation is more short ranged, which is completely consistent
with the result shown in Fig.5.

\begin{table}[ht]
\caption{The spin correlation function between the reference site 26 and all the 10 symmetry inequivalent sites on the 36-site cluster.} 
\centering 
\begin{tabular}{c c c c c c c} 
\hline\hline 
$n$  & &$r$&  &  $J_{r}=0$ & &  $J_{r}=0.15J$ \\ [0.5ex] 
\hline 
27 & &1 & & -0.43838 & & -0.43193 \\ 
15 & &$\sqrt{3}$ & & 0.02314 & &0.13637 \\
14 & &2 & &-0.01802 & & -0.08841 \\
34 & &2 & &0.10547 & & 0.00157 \\
21 & &$\sqrt{7}$ & & -0.00956 & &-0.0058 \\
3 & &3 & &-0.04597 & & -0.02732 \\
32 & &$2\sqrt{3}$ & & 0.01257 & &-0.00959 \\
6 & &$2\sqrt{3}$ & & 0.00636 & &-0.0299 \\
7 & &$\sqrt{13}$ & & -0.01967 & &0.02467 \\
22 & &4 & &0.04443 & & -0.0104 \\[1ex] 
\hline 
\end{tabular}
\label{table:nonlin} 
\end{table}

\begin{figure}[h!]
\includegraphics[width=7cm,angle=0]{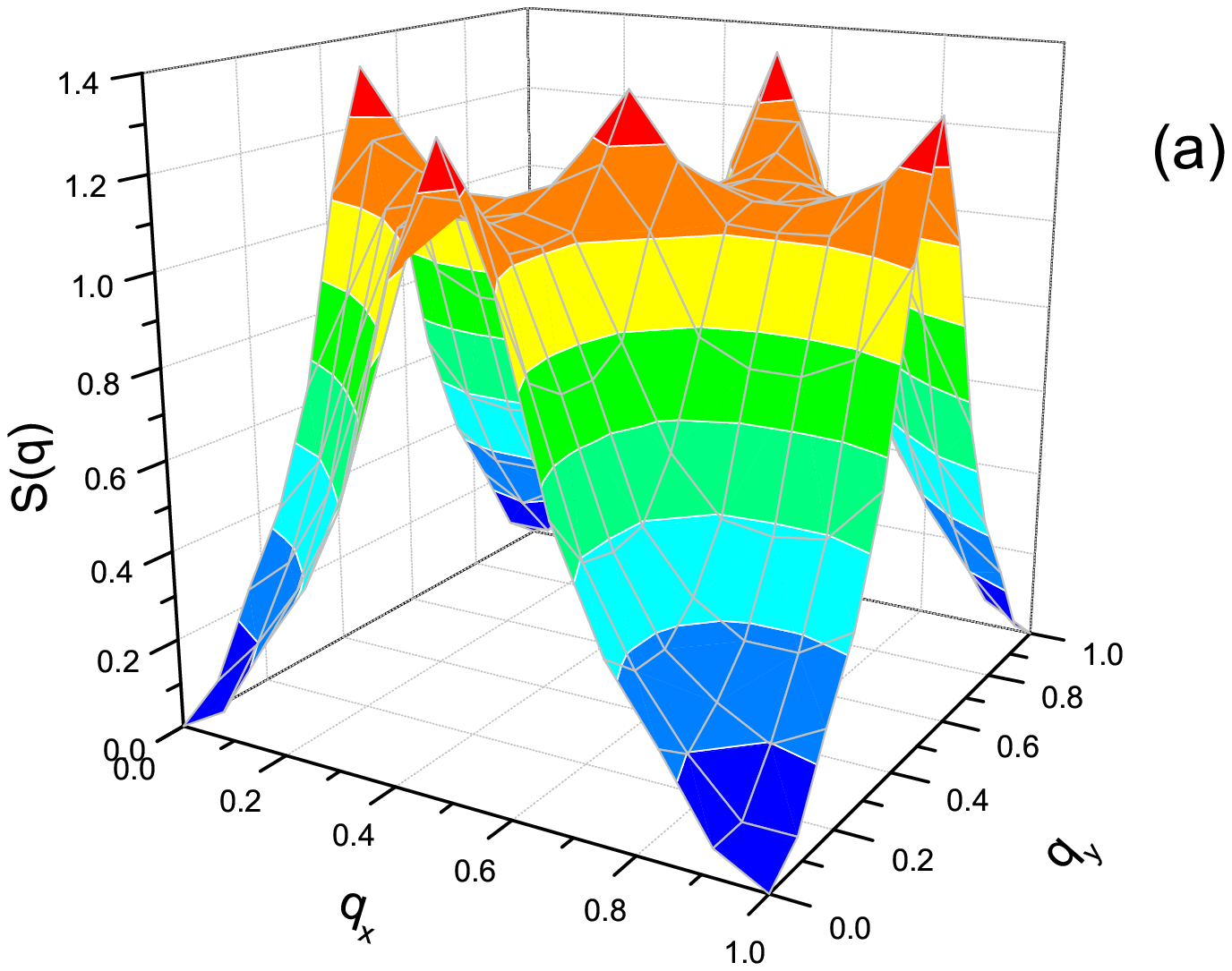}
\includegraphics[width=7cm,angle=0]{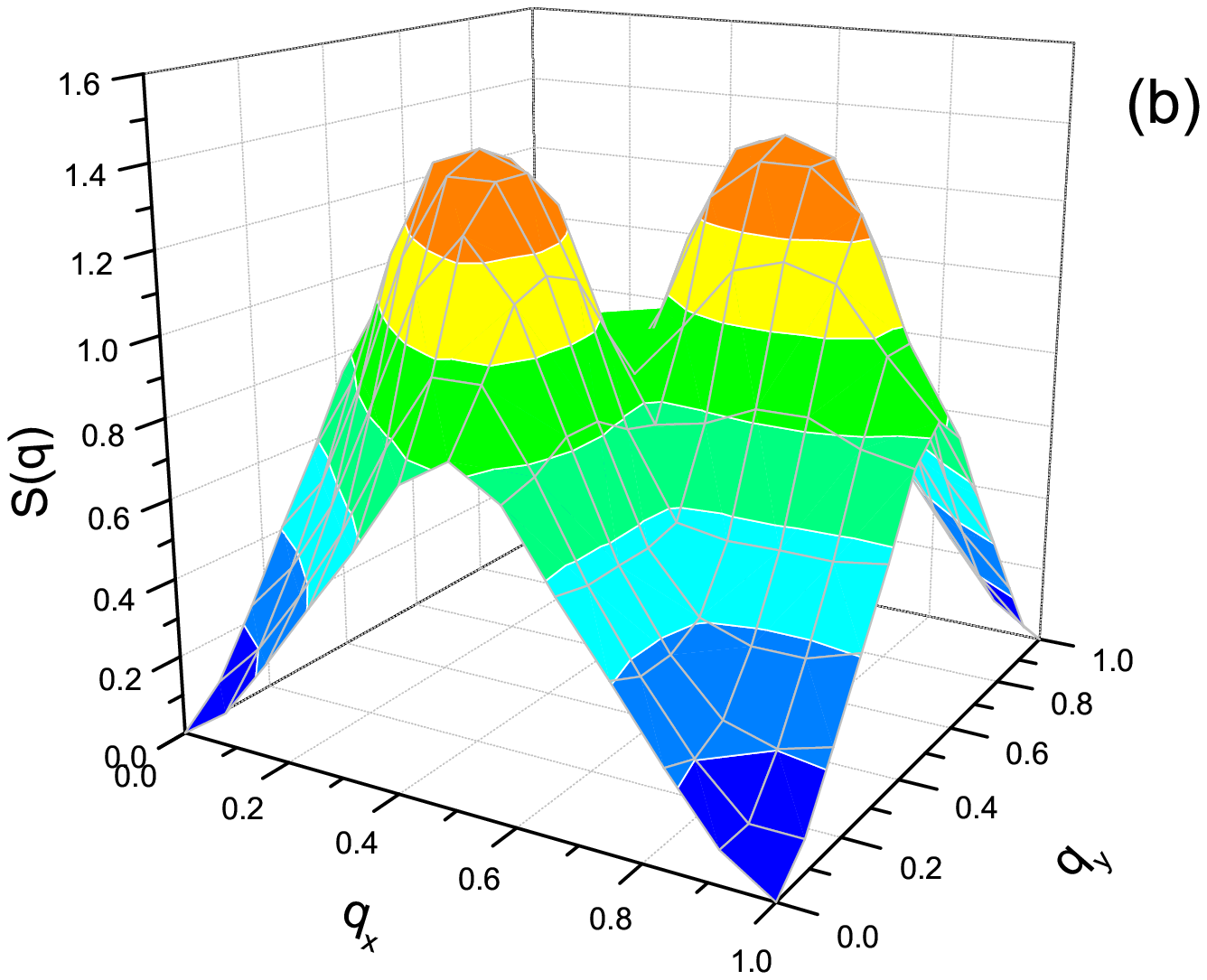}
\caption{The spin structure factor for the (a) pure Heisenberg model
and (b) the model with $J_{r}=0.15J$.} \label{fig5}
\end{figure}

\begin{figure}[h!]
\includegraphics[width=7cm,angle=0]{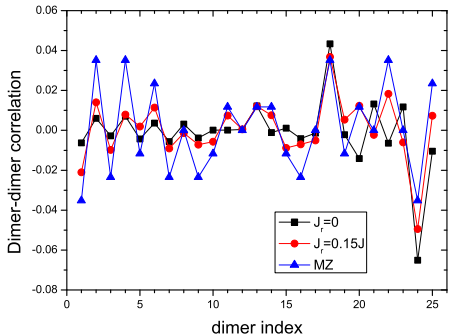}
\caption{The dimer-dimer correlation function between the reference
dimer on bond $(25,26)$ and all the  25 inequivalent dimers on the
36-site cluster. The bonds on which these dimers reside are listed
in Table II. Here $(k,l)$ means that the dimer links site $k$ and
site $l$. MZ denotes the results of the Marston-Zeng spin Pereils
solid state.} \label{fig6}
\end{figure}

\begin{table}[ht]
\caption{The dimer-dimer correlation function between the reference dimer on bond $(25,26)$ and all the  25 inequivalent dimers on the 36-site cluster.
MZ denotes the results of the Marston-Zeng spin Pereils solid state.} 
\centering 
\begin{tabular}{c c c c c c c c c} 
\hline\hline 
$n$  & &$(k,l)$&  &  $J_{r}=0$ & &  $J_{r}=0.15J$ && MZ\\ [0.5ex] 
\hline 
1 & & $(5,6)$ & & -0.00628 & & -0.02105 & & -0.03516 \\
2 & & $(4,5)$ & & 0.00603 & & 0.01403 & &  0.03516 \\
3 & & $(3,4)$ & & -0.00273 & & -0.0098 & & -0.02344 \\
4 & & $(3,8)$ & & 0.0071 & & 0.00793 & & 0.03516 \\
5 & & $(4,8)$ & & -0.0043 & & 0.00196 & & -0.01172 \\
6 & & $(5,9)$ & & 0.00366 & & 0.0114 & &  0.02344 \\
7 & & $(9,14)$ & & -0.00559 & & -0.00909 & &  -0.02344 \\
8 & & $(8,13)$ & & 0.00315 & & -0.00139 & & 0 \\
9 & & $(8,12)$ & & -0.00384 & & -0.00718 & & -0.02344 \\
10 & & $(11,12)$ & & 1.56504E-4 & & -0.00568 & & -0.01172 \\
11 & & $(12,13)$ & & 9.65042E-5 & & 0.00744 & & 0.01172 \\
12 & & $(13,14)$ & & 4.56504E-4 & & 4.99119E-4 & & 0 \\
13 & & $(14,15)$ & & 0.01221 & & 0.01209 & & 0.01172 \\
14 & & $(14,19)$ & & -0.00113 & & 0.00759 & & 0.01172 \\
15 & & $(13,19)$ & & 0.00108 & & -0.0088 & & -0.01172 \\
16 & & $(11,18)$ & & -0.00418 & & -0.00698 & & -0.02344 \\
17 & & $(18,22)$ & & -0.00133 & & -0.00505 & & 0 \\
18 & & $(19,24)$ & & 0.04337 & & 0.03681 & & 0.03516 \\
19 & & $(22,23)$ & & -0.00214 & & 0.00532 & & -0.01172 \\
20 & & $(23,24)$ & & -0.01415 & & 0.01233 & & 0.01172 \\
21 & & $(23,29)$ & & 0.01322 & & -0.00225 & & 0 \\
22 & & $(29,32)$ & & -0.00645 & & 0.01831 & & 0.03516 \\
23 & & $(32,33)$ & & 0.01178 & &  -0.00603 & & 0 \\
24 & & $(34,35)$ & & -0.06509 & & -0.04946 & & -0.03516 \\
25 & & $(1,33)$ & & -0.01045 & & 0.00731 & & 0.02344 \\
[1ex] 
\hline 
\end{tabular}
\label{table:nonlin} 
\end{table}

Now we turn to the dimer-dimer correlation to see if the ring
exchange coupling would drive the system into a state with spin
Pereils type order with broken translational symmetry, which is
hotly debated in the literature. The dimer-dimer correlation
function is defined as follows
\begin{equation}
C(i,j;k,l)=\langle(S_{i}\cdot S_{j})(S_{k}\cdot
S_{l})\rangle-\langle S_{i}\cdot S_{j}\rangle \langle S_{k}\cdot
S_{l}\rangle.
\end{equation}
As in Ref\cite{leung}, we choose the bond between site 25 and 26 as
the reference dimer and calculated the correlation between all the
25 inequivalent dimers with this reference dimer. The results is
tabulated in Table II. As first noticed in Ref\cite{leung}, the
dimer correlation in the pure Heisenberg model modulates at large
distance in close resemblance with a special kind of spin Pereils
solid state(the Marston-Zeng(MZ) spin Pereils solid
state)\cite{marston}, although the amplitude of the modulation is
much weaker. In the MZ spin Pereils solid state, as energy can be
reduced by dimer resonance, the system prefers dimer coverings with
the maximal number of 'perfect hexagon's, on each of which two dimer
configurations(both with three dimers on the hexagon) can resonant
between each other. On the 36-site cluster studied in this paper,
there can be at most two 'perfect hexagon's. As such a state has a
smallest unit cell containing 36 sites, Fourier transform on the
dimer-dimer correlation function on our cluster can not help to
understand such a ordering tendency. We thus compare directly the
modulation pattern in the real space. In Figure 7, we compare the
dimer-dimer correlation of the pure Heisenberg model and the model
with $J_{r}=0.15J$ with the result of the MZ spin Pereils solid
state. As can be seen clearly from the figure, the resemblance
between the modulation pattern with the MZ solid state is greatly
enhanced by the ring exchange coupling. This is in fact not at all
unexpected, as the the ring exchange coupling also encourages
resonance processes around the hexagons. However, as MZ ordering
pattern has a primitive unit cell with 36 sites, our result on the
36-site cluster can not say anything about the long range behavior
of such ordering tendency.

The result presented above can be summarized as follows. First, the
pure Heisenberg model on Kagome lattice is found to be lying exactly
at a quantum critical point and the excitation gap in both spin
channel and the singlet channel should vanish in the thermodynamic
limit. The ring exchange coupling is found to drive the system into
a state with more short ranged spin correlation. Such a spin
disordered state has a local spin correlation pattern in close
resemblance with the antiferromagnetic Heisenberg model on
triangular lattice. At the same time, the state exhibits dimer
correlation in close resemblance with the MZ spin Pereils solid
state, although the long range ordering can not be decided with our
results.

Many issues remains open. In particular, it is interesting to know
if MZ spin Pereils modulation pattern enhanced by the ring exchange
coupling would solidify into a static order in the thermodynamic
limit. This can in principle be answered by methods such as series
expansion calculation. If the modulation remains dynamical, then it
is quite likely that the ring exchange coupling has really drive the
system into a incompressible quantum liquid state, which is long
sought by the researchers in this field. Another issue is about the
nature of the criticality of the pure Heisenberg model, especially,
it is interesting to know if the anomalous singlet dynamics survive
in the thermodynamic limit and how it contribute to the critical
behavior. Finally, we note that as the pure Heisenberg model sit at
a quantum critical point, it is important to take into account the
effect of the ring exchange coupling when compare the experimental
result with theory, as the ring exchange coupling on the hexagons
stands as the most relevant deviation from the Heisenberg limit. It
is interesting to know how the ring exchange coupling will change
the low energy dynamics of the system, especially in the singlet
channel. Our preliminary full spectrum calculation on 18-site
cluster implies that the low energy spectral weight in both the
singlet channel and the triplet channel are reduced with increasing
strength of ring exchange coupling. However, result from larger
cluster and/or from other techniques are obviously needed to settle
down this issue.

The author is grateful to Rong-qiang He for his invaluable help in
the installation and use of the PARPACK library. This work is
supported by NSFC Grant No. 10774187 and National Basic Research
Program of China No. 2007CB925001.


\begin{thebibliography}{9}


\bibitem{review}G. Misguich and C. Lhuillier, in Frustrated Spin Systems, edited
by H. T. Diep (World Scientific, Singapore, 2005); L.Balents, Nature
464, 199 (2010).

\bibitem{elser1} V. Elser, Phys. Rev. Lett. 62, 2405 (1989).

\bibitem{real}J. S. Helton et. al., Phys. Rev. Lett. 98, 107204
(2007); P. Mendels et. al., Phys. Rev. Lett. 98, 077204 (2007); O.
Ofer et. al., cond-mat/0610540.

\bibitem{elser2} C. Zeng and V. Elser, Phys. Rev. B 42, 8436 (1990).


\bibitem{chalker} J. T. Chalker and J. F. Eastmond, Phys. Rev. B 46, 14201 (1992).

\bibitem{singh} R. R. P. Singh and D. A. Huse, Phys. Rev. Lett. 68,
1766 (1992)

\bibitem{leung} P. W. Leung and V. Elser, Phys. Rev. B 47, 5459 (1993).

\bibitem{young} N. Elstner and A. P. Young, Phys. Rev. B 50, 6871 (1994).

\bibitem{lhuillier} P. Lecheminant, B. Bernu, C. Lhuillier, L. Pierre, and P.
Sindzingre, Phys. Rev. B 56, 2521 (1997).

\bibitem{lhuillier3} P. Sindzingre and C. Lhuillier, Europhysics Letters 88,
27009 (2009).


\bibitem{jiang} H. C. Jiang, Z. Y. Weng, and D. N. Sheng, Phys. Rev. Lett.
101, 117203 (2008).

\bibitem{white} S. Yan, D. A. Huse, and S. R. White, Science 332, 1173
(2011).

\bibitem{lauchli2} A. M. L\"{a}uchli, J. Sudan and E.S. S{\o}rensen,
arxiv:1103.1159.

\bibitem{series} R. R. P. Singh and D. A. Huse, Phys. Rev. B 76,
180407(R) (2007). R. R. P. Singh and D. A. Huse, Phys. Rev. B 77,
144415 (2008).

\bibitem{vidal} G. Evenbly and G. Vidal, Phys. Rev. Lett. 104, 187203 (2010).



\bibitem{waldtmann} C.Waldtmann, H.-U. Everts, B. Bernu, C. Lhuillier, P.
Sindzingre, P. Lecheminant, and L. Pierre, Eur. Phys. J. B 2, 501
(1998).

\bibitem{lhuillier2} P. Sindzingre, G. Misguich, C. Lhuillier, B. Bernu, L. Pierre,
Ch. Waldtmann, and H.-U. Everts, Phys. Rev. Lett. 84, 2953 (2000).

\bibitem{impurity} S. Dommange, M. Mambrini, B. Normand and F. Mila, Physical Review B 68, 224416 (2003).

\bibitem{specific}G. Misguich and B. Bernu, Phys. Rev. B, 71, 014417(2005).

\bibitem{lauchli1} A. M. L\"{a}uchli and C. Lhuillier, arxiv:0901.1065

\bibitem{marston} J.B.Marston and C. Zeng, J. Appl. Phys. 69, 5962(1991).

\bibitem{elser3} C. Zeng and V. Elser, Phys. Rev. B 51, 8318 (1995).

\bibitem{mila} F.Mila, Phys. Rev. Lett. 81,2356(1998); M. Mambrini and F. Mila, Eur. Phys. J. B 17, 651
(2000).

\bibitem{auerbach} R. Budnik and A. Auerbach, Phys. Rev. Lett. 93, 187205 (2004).

\bibitem{poilblanc}D. Poilblanc, M. Mambrini, and D. Schwandt, Phys. Rev. B 81,
180402 (2010).

\bibitem{parpack}The PARPACK libary used in this study is downloaded
from http://www.caam.rice.edu/software/ARPACK/.

\bibitem{sachdev}S. Sachdev, Quantum phase transition, Cambridge
University Press 2rd edition, 2011.


\bibitem{hastings} M.B. Hastings, Phys. Rev. B 63, 014413 (2000).

\bibitem{ran}Ying Ran, Michael Hermele, Patrick A. Lee, and Xiao-Gang
Wen, Phys. Rev. Lett. 98, 117205 (2007).

\bibitem{triangular}G. Misguich, C. Lhuillier, B. Bernu and C.
Waldtmann, Phys. Rev. B 60, 1064 (1999).

\bibitem{fermi}O. I. Motrunich, Phys. Rev. B 72, 045105 (2005).

\bibitem{dwave}S. Yunoki and S. Sorella, Phys. Rev. B 74, 014408 (2006); L.
F. Tocchio, A. Parola, C. Gros, and F. Becca, Phys. Rev. B 80,
064419 (2009).

\end{thebibliography}
\end{document}